\documentclass[aps,amsfonts,pra,twocolumn,showpacs]{revtex4}
\usepackage{epsfig,amsmath,amssymb,bm,epsf,graphics,psfrag,verbatim,subfigure}

\def\beq{\begin{equation}}
\def\eeq{\end{equation}}

\def\mass{M}

\def\Dtrans{\tilde{d}}
\def\sbK{K_{{\mathrm{0}}}}
\def\AtomDisp{\varepsilon}

\begin{document}

\title{Emergent superfluid crystals, frustration, and topologically defected states\\
 in multimode cavity~QED}

\author{Sarang Gopalakrishnan$^{1,2}$}

\author{Benjamin L.~Lev$^{1}$}

\author{Paul M.~Goldbart$^{1,2}$}

\affiliation{$^1$Department of Physics and $^2$Institute for Condensed Matter Theory,
University of Illinois at Urbana-Champaign,
1110 West Green Street, Urbana, IL 61801}

\date{March 12, 2009}

\begin{abstract}
We propose that condensed matter phenomena involving the spontaneous emergence and dynamics of crystal lattices can be realized in the setting of ultracold Bose-condensed atoms coupled to multimode cavities. 
Previously, it was shown that in the case of a transversely pumped single-mode cavity, the atoms self-organize at either the even or the odd antinodes of the cavity mode, given sufficient pump intensity, and hence spontaneously break a discrete translational symmetry. 
Here, we demonstrate that in multimode cavities the self-organization brings the additional feature of continuous translational symmetry breaking, via a variant of Brazovskii's transition, thus paving the way for realizations of compliant lattices and associated phenomena, e.g., quantum melting, topological defects, frustration, glassiness, and even supersolidity; such phenomena are absent in ultracold atomic systems when the optical lattices are externally imposed. 
We apply a functional integral approach to this many-body cavity QED system, which enables us, inter alia, to calculate transition thresholds, explore fluctuations near this transition, and determine how such fluctuations are manifest in the light emitted from the cavity.

\end{abstract}

\maketitle

\section{Introduction}\label{sec:intro}

Since the development of modern laser cooling and trapping techniques~\cite{metcalf}, a wide range of phenomena associated with traditional condensed matter physics---ranging from Mott insulators~\cite{jaksch, bloch:mott} to Tonks-Girardeau gases~\cite{bloch:tonks}---have been realized in ultracold atomic systems. The value of these realizations stems from their exquisite tunability and purity.  A shortcoming of traditional optical lattice settings, however, lies in the artificial nature of the lattice itself: its potential wells are fixed by external lasers, rather than arising spontaneously from many-body effects, which generate spatial structure in liquid crystals, solids, and even glasses.  Therefore, several areas of condensed matter research---e.g., soft condensed matter physics~\cite{deg:nobel}, as well as phenomena such as supersolidity~\cite{supersolids}---have remained inaccessible to the physics of ultracold atoms in artificial optical lattices, because of the absence of an emergent, compliant lattice, capable of exhibiting, e.g., dynamics, defects, and melting.  Tunable ultracold-atom versions of these condensed matter situations are especially welcome, as several fundamental issues remain unresolved, such as the dynamics of glassy media.

A major advance toward self-generated, dynamical optical lattices was the discovery of cavity-induced self-organization~\cite{ritschprl, ritschpra, Lewenstein06}, which may be explained as follows. Consider $N$ two-level atoms in a \textit{single-mode} optical cavity, interacting with the cavity mode and a pump laser oriented transverse to the cavity axis. The atoms  coherently scatter light between the pump and cavity modes. Atoms arranged at every other antinode of the cavity field (i.e., one wavelength $\lambda$ apart) emit in phase and populate the cavity with photons, leading to the collective, superradiant enhancement of the atom-cavity coupling by a factor proportional to the number of organized atoms.  If the pump laser is of sufficient intensity and is red-detuned from the atomic transition (so that the atoms are high-field seekers),  an instability arises: the superradiance makes atoms couple more strongly to the cavity mode, thus trapping themselves at either the even or the odd set of antinodes, which leads to greater superradiance, stronger trapping, and so on. The system reaches a spatially modulated steady state when the energetic gain from superradiance is balanced by the cost, in kinetic energy or repulsive interactions, of confining the atoms to the even or odd sites of the emergent lattice.

Signatures of this symmetry-breaking, nonequilibrium phase transition have been observed experimentally~\cite{vuletic:prl}, and the resulting cavity-assisted laser cooling has been suggested as a method for cooling non-two level atomic or molecular systems~\cite{chu-2000:prl}.  Although this transition has been treated in a mean-field approximation for a single-mode cavity, the threshold for this transition has not been calculated beyond mean-field theory, which is crucial for analyzing the technique's utility for molecular laser cooling---a potentially ground-breaking prospect~\cite{lev}.

Our interest in atomic systems confined in multimode cavities stems from the vast family of possibilities for ordering and fluctuations that they can sustain.  Single-mode cavities, unlike externally imposed optical lattices, feature a lattice \textit{amplitude\/} that is emergent and a discrete symmetry breaking between even and odd antinode ordering. However, the locations of the lattice antinodes are not themselves emergent, instead being fixed by the cavity geometry.  By contrast, even in the simplest multimode cavity (i.e., the ring cavity, which supports two counterpropagating modes) the atoms collectively fix the locations of the antinodes, thus spontaneously breaking a continuous translational symmetry.  As in real crystals, this results in an emergent rigidity with respect to lattice deformations.  In cavities possessing a family of degenerate modes, one can anticipate realizing further phenomena associated with crystallizing systems, including topological defects such as dislocations and domain walls.

\begin{figure}
	\centering
		\includegraphics{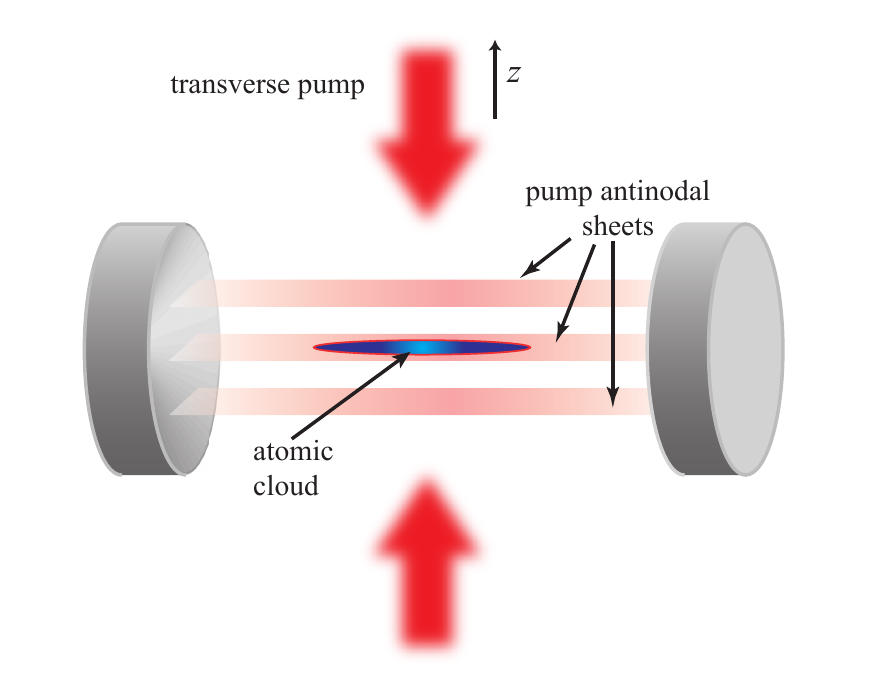}
	\caption{
	\textbf{The layered atom-cavity system.} 
	The cavity is transversely pumped by two counterpropagating lasers, which create a deep optical lattice, and confine atoms at the antinodes of this lattice. In the depicted situation, the atoms occupy a single equatorial antinodal layer.}
	\label{fig:cav3d2}
\end{figure}

The purpose of this paper is to develop a field-theoretical framework for exploring the quantum statistical mechanics of correlated many-atom, many-photon systems confined in multimode optical cavities.  This framework enables us to treat phenomena such as quantum phase transitions exhibited by the atom-cavity system and, in particular, to analyze the consequences of collective fluctuations, which play a pivotal role in the determination, formation and slow dynamics of various ordered states.  We apply our framework to the case of a quasi-2D cloud of atoms confined in a concentric optical cavity, and find a transition to a spatially modulated state that realizes Brazovskii's transition~\cite{brazovskii}.  The transition persists to zero temperature, thus becoming a quantum phase transition of an unusual universality class.  
In addition, we address the imprint of the associated quantum fluctuations in the correlations of the light emitted from the cavity.
Finally, we generalize our model to incorporate strongly layered 3D distributions of atoms, and find that (in certain regimes) such systems are unable to order globally because of frustration, and are expected, therefore, to develop inhomogeneous domain structures. Indications of such phenomena were observed in quantum Monte Carlo simulations~\cite{ritsch:personal}. In this work we identify the origin of these domains.

The states we consider are \lq\lq supersolids\rlap,\rq\rq\ in the sense that they are characterized by off-diagonal long-range order (arising from the Bose-Einstein condensation of the atoms) as well as emergent crystalline (diagonal) order.  A key difference between our system and those studied in Refs.~\cite{lew02,sun07} is that the solidity is associated with a broken continuous symmetry, and can therefore be used to perform ultracold-atomic versions of experiments studying the transport properties of supersolids~\cite{supersolids}.

To date, a key theme in AMO research has been the vision of using atoms to simulate the quantum dynamics of electrons propagating through static (i.e., ``hard'') lattices. One of the aims of this paper is to suggest that the ultracold atomic physics setting of cavity QED may be brought to bear on systems that are ``soft," possessing spatial structures that are readily deformed by stresses and fluctuations---a possibility that is viable precisely because the lattice itself is a dynamical entity.

\section{Elements of the model}

Near the ordering threshold, the strongest potential acting on the atoms is due to the standing wave of the pump lasers along the $z$ direction (see Fig.~\ref{fig:cav3d2}). The strength of the pump lasers at threshold determines the extent to which an atom is confined to an antinodal sheet of the pump standing wave.  The weaker the coupling $g$ between a single atom and a cavity mode, the stronger the laser must be to achieve threshold; in the regime we consider, $g$ is weak enough that the atomic distribution is layered at threshold.  Therefore, as a first step we analyze a system of atoms confined to a single layer.  Such a situation can be realized via selective loading techniques~\cite{dalibard}.

We start with the Hamiltonian $H$ for a two-dimensional system of $N$ two-level atoms interacting with the degenerate cavity modes and a spatially uniform transverse pump laser field~\cite{walls}:
\begin{eqnarray}
H & = &
\sum_{n = 1}^{N}
\left[ \frac{\mathbf{p}_n^2}{2\mass} + V(\mathbf{x}_n) + \omega_A \sigma^z_n
\right]+
\sum_{\mu = 1} \omega_C \,a_\mu^\dagger\, a_\mu^{\phantom{\dagger}} \\
& + &
i \sum_{n,\mu} (g_\mu(x_n) \, a_\mu^\dagger \, \sigma_n^- - h.c.) +
i \Omega \sum_n (\sigma_{n}^{-} - \sigma_n^+) +
H^{\prime}.
 \nonumber
\end{eqnarray}
The $\{a_\mu^{\phantom{\dagger}}\}$ are cavity photon annihilation operators;
$g_\mu(x) \equiv g\,\Xi_\mu(x)$,
where $g$ is the atom-cavity coupling and $\Xi_\mu$ is the normalized mode function of mode $\mu$;
$\Omega$ is the pump laser strength;
and the $\sigma$ operators are Pauli matrices that act on the pseudospin of the two-level atom.
$H^{\prime}$ consists of dissipative terms arising from cavity loss ($\sim\kappa$) and spontaneous decay ($\sim \gamma$);
the effects of these terms will be discussed later.
We assume throughout that the pump laser is slightly red-detuned from the cavity mode,
and that both are substantially red-detuned from the atomic transition,
so that $\omega_L \gg (\omega_A - \omega_L) \gg (\omega_C - \omega_L) > 0$.
For compactness, we re-express the frequencies in terms of detunings
$\Delta_A \equiv \omega_A - \omega_C \approx \omega_A - \omega_L$ and
$\Delta_C \equiv \omega_C - \omega_L$. Furthermore, we assume that the atom-cavity coupling $g$ is weak enough that 
$g^2 N/\Delta_A \ll \kappa, \Delta_C$, where $\kappa$ is the linewidth of the cavity.

\section{Ring cavity}

Let us first consider the ring cavity, which sustains two degenerate counter-propagating cavity modes of mode functions $e^{\pm iKx}$ and wavevectors $\pm K$~\cite{siegman}.  If the atoms scatter pump photons into both modes, and do so with a fixed phase relation between the modes, a standing wave is created in the cavity, and the system undergoes the same type of instability as in the single-mode case.  There is, however, a crucial difference: the ring cavity has a continuous symmetry, because the cavity geometry does not fix the location of the antinodes.  Instead, the atoms must collectively fix a set of points, spaced $\lambda$ apart, at which to crystallize.  The ordered state in the ring cavity is the simplest case, in ultracold-atom physics, of a system that spontaneously breaks a full, continuous, translational symmetry of space.

In the mean-field approximation, the cavity mode amplitudes obey the classical equations of motion arising from $H$:
\begin{eqnarray}
&& a_\pm e^{-i\omega_L t} =
\frac{\Omega g/\Delta_A }{\Delta_C + i \kappa + g^2 N/\Delta_A} \int N(x)\, e^{\pm iKx} dx
\nonumber
\\
&& \qquad + \! 
\frac{g^2 a_\mp/\Delta_A }{\Delta_C + i \kappa + g^2 N/\Delta_A}\int N(x)\, e^{\mp 2iKx} dx\,.
\label{eq:modetime}
\end{eqnarray}
Near threshold, and in the weak-coupling limit (which implies a large threshold for the pump), we can drop the second term on the RHS of Eqs.~(\ref{eq:modetime}), in which case the equations for $a_+$ and $a_-$ decouple.
\def\HatomMF{\widetilde H}
We now eliminate the cavity modes from $H$, using the solutions of Eqs.~(\ref{eq:modetime}), second-quantize the atoms as bosons, and expand the atomic field operators about the condensate as $\sqrt{N}+\psi_K$, to obtain for the approximate atomic Hamiltonian:
\beq
\HatomMF = \frac{\hbar^2 K^2}{2\mass}
\psi_K^\dagger \psi_K^{\phantom{\dagger}} -
\frac{\zeta}{2} \Big[2 \psi_K^\dagger \psi_K^{\phantom{\dagger}} +
\psi_K^{\phantom{\dagger}} \psi_{-K}^{\phantom{\dagger}} +
\psi^\dagger_K \psi^\dagger_{-K}\Big],
\eeq
where $\zeta N \equiv \frac{g^2 \Omega^2 \Delta_C}{\Delta_A^2 \sqrt{\Delta_C^2 + \kappa^2}}$.
Performing a Bogoliubov transformation~\cite{ll9} yields the result that the polariton excitation of momentum $K$ has energy
\beq
\epsilon_K = \sqrt{\frac{\hbar^2 K^2}{2\mass}} \sqrt{\frac{\hbar^2 K^2}{2\mass} - 2 \zeta N}.
\eeq
[For low nonzero temperatures $T$ this dispersion relation continues to hold, provided the particle number $N$ is replaced by the condensate fraction $N_0(T)$.]\thinspace\ Upon increasing $\zeta$ (e.g., by increasing $\Omega$) the uniform condensate becomes unstable when $\epsilon_K = 0$, i.e., when
\beq
\frac{\hbar^2 K^2}{2\mass} = 2 \hbar^2 \frac{g^2 \Omega^2 \Delta_C}{\Delta_A^2 (\Delta_C^2 + \kappa^2)} N.
\eeq
For laser strengths greater than threshold, the uniform condensate is unstable with respect to a state that is spatially modulated with wavelength $\lambda$, as discussed in Sec.~\ref{sec:intro}. This spatial modulation is due to the macroscopic occupation by atoms of modes $\psi_{\pm K}$. The order parameter for the transition is $\langle \psi_K \rangle$, a complex number whose phase, $\phi$, determines the position of the antinodes, and which represents the Goldstone mode associated with the spontaneously broken translational symmetry. To find the steady-state value of the order parameter, we enforce number conservation,
$\psi^\dagger_K\, \psi_K^{\phantom{\dagger}} +
\psi^\dagger_{-K}\, \psi_{-K}^{\phantom{\dagger}} +
\psi^\dagger_0 \, \psi_0^{\phantom{\dagger}} = N$,
and minimize the energy with respect to the order parameter (note that $\zeta N \sim g^2 N$ is an intensive quantity):
\beq
\langle\psi_K^{\phantom{\dagger}}\rangle =
e^{i\phi}
\sqrt{\frac{2 \zeta N - \frac{\hbar^2 K^2}{2\mass}}{4 \zeta N}N}\, .
\eeq
Because $\langle\psi_K^{\phantom{\dagger}}\rangle\rightarrow 0$ at threshold, the transition is continuous. These results complement those of Nagy \textit{et al}.~\cite{domokos08}, who arrived at similar results via the Gross-Pitaevskii equation. They are slightly more general as they hold for nonzero temperatures below $T_c$.

\section{Concentric cavity}

For simplicity of presentation, most of this section will focus on one particular kind of multimode optical cavity, the concentric cavity. A concentric cavity consists of two (incomplete) spherical mirrors with coincident centers of curvature.  It can be thought of as a single sphere with a large strip around the meridian removed.  Perfectly concentric cavities are unstable---they do not support localized optical modes---but it is possible to make nearly concentric cavities having stable modes that are still frequency-degenerate to within the cavity's linewidth. The analysis below focuses on concentric cavities because of their familiar symmetry structure, but almost all of it applies to \textit{any} cavity having a family of degenerate modes, and in particular to the confocal cavity.

\subsection{Effective action}

The mean-field theory outlined above for the ring cavity is not adequate for describing ordering in concentric or confocal cavities, because the larger set of possibilities for ordering increases the importance of fluctuations. To incorporate fluctuations, it is convenient to formulate the description in terms of functional integration.  This enables us to integrate out the atomic excited states and all of the photon states so as to construct an effective description that focuses on the motion of the (ground state) atoms. 
In general, the atom-cavity system is driven, owing to the pump laser, and dissipative, owing to cavity loss and spontaneous emission. Consequently, there is an energy flux through the system, which is proportional to $N g^2$ below threshold and to $ \langle \Xi_\mu \rangle N^2 g^2$ above threshold, where $\langle \Xi_\mu \rangle$ is the order parameter for ordering in mode $\mu$. As $g \propto V^{-1/2}$, $V$ being the volume of the system, in order to arrive at the thermodynamic limit at constant density, one keeps $g^2 N$ constant (and small relative to $\kappa$, as mentioned in Sec.~II) as $g$ decreases and $N$ increases~\cite{ritschpra}. Under these assumptions, the energy flux per particle through the system is small near threshold, and we can approximate the system as dissipative but undriven.
The dissipation can then be treated in the manner of Caldeira and Leggett~\cite{cl}. 
Next, we make the rotating-wave approximation, which transforms fields that rotate at $\omega_L$ in the complex plane to static fields, so that the drive becomes a static coupling between the atoms and the cavity. Thus, we arrive at an effective description of the atom-cavity system in terms of equilibrium quantum statistical mechanics.

In the uniform phase (i.e., below threshold), $\Omega^2 \gg g^2 \langle |a_m|^2 \rangle$, and the leading part of the effective atomic action is:

\def\mats{\nu}

\begin{eqnarray}
&& S_0  = 
\sum_\mats d^2 x\,
\psi^\dagger(\omega_\mats,\mathbf{x})
(i \omega_\nu - \nabla^2 - \mu)\,
\psi(\omega_\mats, \mathbf{x}) \\
&& \quad - 
\frac{\zeta}{N}\sum_{\mats,n_i}
\int d^2 x\,
g_\mu (\mathbf{x})\,
\psi^\ast(\omega_{\mats_1},\mathbf{x})\,
\psi     (\omega_{\mats_2},\mathbf{x})
\nonumber \\
&& \quad \times \!
\int \! d^2x'
g_\mu(\mathbf{x}')
\psi^\ast(\omega_{\mats_3},\mathbf{x}')
\psi(\omega_{\mats_4},\mathbf{x}')\,
\delta_{\mats_1 + \mats_2,\mats_3 + \mats_4},\nonumber
\end{eqnarray}
where $\{ \omega_\mats \}$ are Matsubara frequencies~\cite{ll9}. The second term may be understood as follows: if the system were in a cubic box, so that the mode functions were $e^{i\mathbf{\sbK\cdot x}}$, this term would have the form $\rho_{\mathbf{\sbK}}\, \rho_{\mathbf{-\sbK}}$, where $\rho_{\mathbf{\sbK}} = \langle e^{i \mathbf{\sbK\cdot x}} \rangle$ is a Fourier component of the density. Such a term is commonplace in the theory of crystallization (see, e.g., Ref.~\cite{mctague}), in which $\rho_{\mathbf{\sbK}}$ is often defined to be the order parameter. If the uniform state is a condensate, $\rho_{\mathbf{\sbK}} = \sum_{\mathbf{q}} \langle \psi^\dagger_{\mathbf{q}}\, \psi^{\phantom{\dagger}}_{\mathbf{\sbK}} \rangle \approx \sqrt{N_0} \langle \psi_{\mathbf{\sbK}} \rangle$, i.e., the expectation value approximately factorizes owing to the presence of off-diagonal long-range order. Therefore, $\langle \psi_{\mathbf{\sbK}} \rangle$ is a legitimate order parameter for crystallization, as we implicitly assumed in the previous section.

For the concentric cavity, the mode structure for the light is referenced by the index in the pump direction $l$, the angular index $m$, and the radial index $n$; along the equatorial plane, $l = 0$ modes have the highest amplitude, and we shall restrict to $l = 0$ in this section. In this case, the mode structure for both the light and the atoms is indexed by $m$ and $n$. Assuming that the relevant cavity modes have spatial structure on lengthscales far smaller than the cavity size, the interaction term takes the form $\rho_{mn}\, \rho_{-mn}$, where $m+n \simeq \sbK R$, with $R$ being the radius of the cavity and $\sbK = 2\pi/\lambda$. As the cavity-mediated interaction draws atoms into antinodal wells one wavelength apart from one another, it generates a softening in the atomic dispersion relation $\AtomDisp_{mn}$, i.e., a trough near $m+n \simeq \sbK R$. We now develop a low-energy effective action for the atomic degrees of freedom in the vicinity of this trough, in order to examine whether it becomes energetically favorable, relative to the spatially uniform state, for the system to order. We consider two cases: (i)~the classical transition at $T > 0$ in the absence of a condensate, for which $N_{\mathbf{\sbK}}$ does not factorize, and (ii)~the quantum phase transition at $T = 0$, in the presence of a condensate, near which the dominant fluctuations are quantum mechanical. 

In the classical case, we introduce collective coordinates $\rho_{mn}$ to represent atomic density fluctuations, and then integrate out the microscopic atomic degrees of freedom, to arrive at the effective action in terms of rescaled dimensionless fields:

\begin{eqnarray}
&& S_{\mathrm{eff}} = \sum_{mn} \left[ \tau + \chi (m + n - \sbK R)^2 \right] \rho_{mn} \rho_{-mn} \\
&& + \frac{N \zeta^2 (k_B T)^2}{\hbar^6 \sbK^4 \rho_0/8\mass^3}\sum_{m_i, n_i} \! \rho_{m_1n_1} \rho_{m_2n_2} \rho_{m_3n_3} \rho_{m_4n_4} \, \delta_{\sum m_i} \delta_{\sum n_i}, \nonumber
\end{eqnarray}
where $\rho_0$ is the area density of atoms, 

\beq
\tau = 1 - \frac{N \zeta k_B T}{\hbar^4 \rho_0 \sbK^2 / 4\mass^2},
\eeq
and $\chi$ is a parameter that describes the width of the trough at $\sbK$. Contributions to $\chi$ arise from the intrinsic linewidth of the cavity modes and from the further broadening due to the atoms-cavity coupling. There is no cubic term in the action because, to meet the requirements for \lq\lq momentum\rq\rq\ conservation, such a term would have to involve at least one mode with $m \geq \sbK R / 2$. However, modes with most of their structure in the angular direction are suppressed because (i)~such modes tend to have higher diffractive losses~\cite{siegman}, and (ii)~the atomic cloud is confined to the intersection of pump and cavity modes, near the center of the cavity, and is therefore located mostly near modes with low angular momentum, i.e., $m \approx 0$.

Consequently, the effective action is not---as one might have expected---the Landau free energy for crystallization, but is instead a variant of Brazovskii's free energy~\cite{brazovskii}, which describes phase transitions from isotropic to striped structures in various condensed matter settings, ranging from diblock copolymers~\cite{leibler} to convective patterns~\cite{swift:rbc}.  Brazovskii's phase transition differs from many commonly studied phase transitions in an important respect---the low-energy fluctuations active near most phase transitions are clustered about a point, or isolated set of points, in momentum space (frequently the origin).  By contrast, for the Brazovskii case, as long as dimension $d\geq 2$, the low-energy fluctuations are clustered about a circular shell of nonzero radius, given by the stripe wavevector.  This has two important consequences.  
First, the low-energy density of states is effectively one-dimensional, and scales as $(\epsilon - \epsilon_0)^{-1/2}$, where $\epsilon_0$ is the energy of an excitation at wavevector $\sbK$, regardless of the physical dimensionality of the system; therefore, the physics of Brazovskii's model is independent of the physical dimensionality of the system, as long as $d \geq 2$.  Second, the phase space for fluctuations is larger; therefore, fluctuations are strong enough to control not only the critical exponents at the transition but even the order of the transition itself. 

The mean-field threshold $\Omega_{\mathrm{mf}}$ can be read off from the action $S_{\mathrm{eff}}$ by setting $\tau = 0$:
\beq
\frac{\hbar^2 \sbK^2}{2M} \frac{\hbar^2 \rho_0}{2\mass\! N} \ln \left(\frac{\hbar^2 \sbK^2}{2 \mass} \frac{1}{\mu} \right) = \frac{\hbar \Delta_C \, g^2 \, \Omega_{\mathrm{mf}}^2}{\Delta_A^2\, (\Delta_C^2 + \kappa^2)} k_B T,
\eeq
and the transition appears to be continuous. (Note that, although $T > 0$, we have assumed that $T$ is low enough that the thermal de Broglie wavelength exceeds the lattice spacing; therefore, this result does not apply to point particles.) However, the true threshold for ordering, $\Omega_{\mathrm{th}}$, once fluctuations are accounted for, is higher:

\beq
\Omega_{\mathrm{th}}^2 - \Omega_{\mathrm{mf}}^2  \sim \left[ \frac{\Delta_A^2 (\Delta_C^2 + \kappa^2)}{g^2 \Delta_C} \right]^{-1/3} \left[ \frac{\Omega_\mathrm{th}^4 \sqrt{k_BT}}{\mu} \right]^{2/3} \chi^{1/3}.
\eeq
In addition, the transition is driven first-order: the order parameter jumps discontinuously from zero to a nonzero value. As with other first-order transitions, Brazovskii's transition proceeds by the nucleation and growth of droplets of the stable phase.  The morphology of these droplets is known to be unusually rich: there are regimes dominated by anisotropic and diffuse droplets, as well as ones in which the droplets form ``focal conic'' structures~\cite{swift:bubbles}.  Qualitatively similar phenomena are expected to occur in the cavity QED setting, although the nature of the crossovers between regimes may be different.

\begin{figure}
	\centering
		\includegraphics{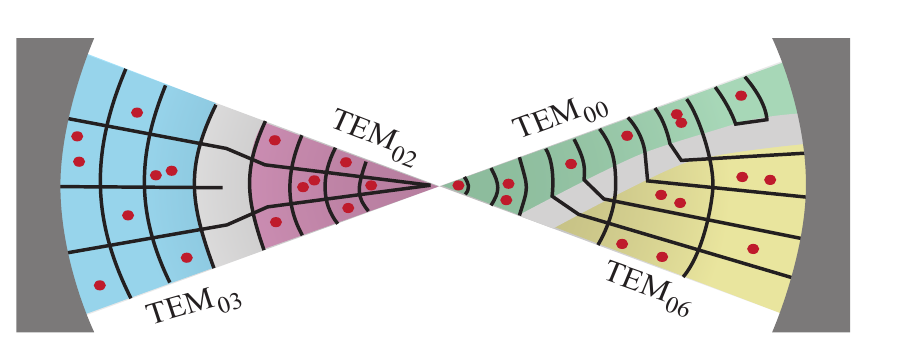}
	\caption{\textbf{Ordered state with defects.} The ordered states form two-dimensional patterns. This sketch shows a regime near threshold, with domains locally populating distinct cavity modes at the equatorial plane.  
	Domains can be punctuated by dislocations (shown in the left-half of the figure), but might also show textural variation in space (right-half of figure). The black lines represent nodes of the cavity field, which separate ``even'' and ``odd'' antinodes. The atomic population per site is not fixed because the atoms are Bose condensed.}
	\label{fig:ordstpic}
\end{figure}

Having addressed the classical regime of the interacting Bose gas, we now turn to the other limit of interest: the quantum, zero-temperature limit. In this regime, an argument adapted from Ref.~\cite{ll9} yields the effective low-frequency action
\begin{eqnarray}
S & = & \int d\omega\, 
\sum_{mn}
\frac{1}{\zeta} \left[
\tau + \omega^2 + \chi' (m + n - \sbK R)^2
\right]
|B_{\omega mn}|^2 \nonumber \\
& & + \, U \int dt \, d^2x \, |B(\mathbf{x},t)|^4 + \ldots,
\end{eqnarray}
where we have augmented the action with a contact repulsion term associated with the energy scale $U$. The field $B$ describes the Bogoliubov quasi-particles, $\tau' = (\hbar^2 \sbK^2/ 2\mass)^2 - 2 \zeta \hbar^2 \sbK^2/ 2 \mass$ and $\chi'$ serves, as $\chi$ did in the classical case, as a broadening of the constraint $m+n\simeq\sbK R$. The mean-field threshold in this case, which occurs when $\tau = 0$, is the same as in the ring cavity, discussed in Sec.~III. The extra dimension arising from the integral over $\omega$ changes the spectrum of fluctuations; instead of a thin ribbon of fluctuations, we must now consider an anisotropic tube. In the limit that temporal fluctuations are of much lower energy than spatial ones, Brazovskii's analysis is applicable; in general, however, we must adapt Brazovskii's argument to the case of $\Dtrans$ (rather than one) dimensions transverse to the line of soft modes $m+n\simeq \sbK R$. In this case, power counting suggests that the case of $\Dtrans=2$ forms a boundary between transitions that are driven first-order and those that are not. Related issues have recently been addressed by Ref.~\cite{shankar}.  It is plausible that at $\Dtrans=2$, the variation of some parameter could tune the system from a Brazovskii-like regime in which the quantum phase transition is first-order, via a tri-critical point, to a regime in which the transition is continuous.

\subsection{Correlation functions of emitted light}

Our approach can also be used to compute atom-atom correlation functions via standard functional or diagrammatic techniques.  Such atomic correlations are experimentally obtainable through post-processing of light scattering and absorption images~\cite{Demler04}. Additionally, transmission through the cavity mirrors provides an experimentally accessible, real-time diagnostic channel for critical fluctuations in the correlations of the light emitted from the cavity. These optical correlations can be computed from our effective atomic theory by the field-theoretic technique of introducing a formal source term $h_\mu a_\mu$ that couples linearly to the cavity mode $\mu$.  Near threshold, one expects the occurrence of \lq\lq critical slowing down\rlap,\rq\rq\ which manifests itself through the development of long-time (with respect to $\kappa$) correlations in the emitted light, possibly characterized by an anomalous power-law. 

For instance, the low-frequency temporal photon correlation functions just below threshold are related to atomic correlation functions as follows:

\begin{eqnarray}
&& \langle a^\dagger_{mn}(\omega) a_{mn}(\omega) \rangle \!= \! \frac{\zeta}{\hbar\Delta_C}
 \!\int\! d\omega'd\omega''d^2xd^2x' g_{mn}(x) g_{mn}(x') \nonumber\\
&& \quad \times \langle \psi^\dagger(\omega',x) \psi(\omega - \omega',x) \psi^\dagger(\omega'',x') \psi(\omega - \omega'',x') \rangle .
\end{eqnarray}
The utility of this result is that it enables us to translate scaling arguments for the atomic correlations into scaling arguments for the photon correlations. At the Gaussian level, the photon correlators diverge as the control parameter $\tau' \rightarrow 0$ as follows:

\beq
\langle a^\dagger_{mn}(\omega) a_{mn}(\omega) \rangle \sim
\frac{1}{\omega^2 - \tau' + i 0}.
\eeq
However, this result is expected to be inaccurate very close to the phase transition, as a result of nonlinear interactions between fluctuations. The theory of the divergence of correlation lengths and times near a quantum phase transition is described in Refs.~\cite{sondhi,sachdev}.  It is believed that the chief consequence of an energy flux near a quantum phase transition is to act as an effective  temperature~\cite{mitra1,mitra2,sondhi}; if one assumes that this holds for the concentric cavity, one can adapt the finite-size scaling hypothesis to predict that 

\beq
\langle a^\dagger_{mn}(\omega) a_{mn}(\omega) \rangle \sim
\gamma^{-\eta} f(\omega \gamma^{-z/2}, \xi \gamma^{1/2}, \kappa / \gamma),
\eeq
where $f$ is a scaling function, $\xi \sim \big((\hbar^2K^2/2\mass) - 2 \zeta\big)^{-\nu}$ is the spatial correlation length of the system, $\nu$ is a critical exponent, and $\eta$ is the scaling dimension of the correlation function.

\begin{figure}
	\centering
		\includegraphics{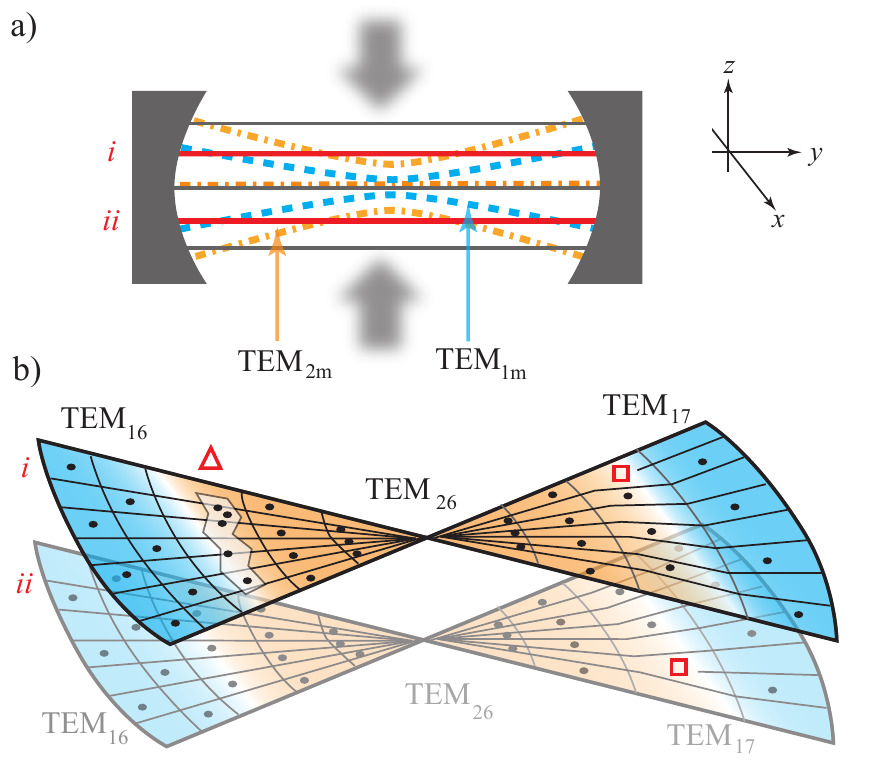}
	\caption{\textbf{Effects due to frustration.} Atoms are loaded into sheets~(i) and (ii), marked in thick red lines in panel~(a), which are an integer number of pump wavelengths apart. 
The blue [dashed] and orange [dash-dot] curves are, respectively, antinodal regions of the modes $\mathrm{TEM}_{1m}$, which has low intensity near the centers of sheets~(i) and (ii), and $\mathrm{TEM}_{2m}$, which has low intensity away from the centers of sheets~(i) and (ii).  Near the center of each sheet, atoms crystallize into $\mathrm{TEM}_{2m}$; away from the center, they crystallize into $\mathrm{TEM}_{1m'}$.  Within a sheet, regions may be separated by a discommensuration, e.g., in the left side of panel (b), or a dislocation, e.g., in the right side of panel~(b).  Between sheets, the opposite parity of adjacent modes leads to frustration, which precludes ordering, as in the regions indicated by a $\triangle$ and a $\Box$.  Dislocations ($\Box$) are less energetically costly than discommensurations ($\triangle$) because they are more localized.}
\label{fig:frustra}
\end{figure}

\subsection{Character of ordered states and their defects}

Soft condensed matter systems that undergo Brazovskii's transition commonly form lamellar patterns. The present realization does not, owing to the influence of boundary conditions on the optical mode structure, and therefore on the possibilities for atomic crystallization. Instead, in the concentric cavity, the ordered states follow the optical mode patterns, which may be visualized as distorted checkerboard patterns, as shown in Fig.~\ref{fig:ordstpic}. The atomic density locally selects amongst the cavity modes by crystallizing into the \lq\lq black or white squares\rq\rq\ of the selected mode. Domains of crystallization are separated by localized defects or extended textures, which are expected to coarsen over time. In physical realizations, the states of crystallization are not exactly degenerate, as optical modes having more angular structure (i.e., larger $m$) are of lower finesse; on the other hand, repulsive interactions have a stronger impact on atoms crystallizing into modes of lower $m$. Such effects can be accounted for within our model by the introduction of external fields that bias the system towards crystallizing into certain modes.

\section{Other multimode cavities}

The best-known example of a multimode cavity is the confocal cavity, in which all the even TEM modes are degenerate~\cite{siegman}.  (It is also possible to make multimode cavities in which every $p^{\rm th}$ TEM mode is degenerate.)\thinspace\  These cavities have the practical advantage over the concentric cavity that their stability criteria are easier to fulfill.  To the extent that it is legitimate to think of such cavities as having a continuous family of degenerate modes (i.e., if they possess a large number of modes that are both degenerate and not heavily suppressed by diffractive losses), the self-organization transition in these cavities should belong to Brazovskii's universality class, and our analysis of the transition itself extends to these models. Where confocal cavities differ from concentric ones is in the geometry of the ordered states and of their defects, which is much more involved in the confocal case because of its less-evident symmetry structure.

\section{Frustration in layered systems}

We have discussed how an equilibrium atomic cloud confined by the pump laser to a plane near the equatorial plane of the cavity spontaneously crystallizes globally into one of a family of degenerate checkerboard arrangements. Now let us consider an atomic cloud confined to a plane \textit{away} from the equator of the cavity. In this case, spontaneous crystallization still occurs, but, as we shall now explain, the precise arrangement into which the atoms crystallize varies across the plane. This is a consequence of frustration (see, for example, Ref.~\cite{parisi}, Part~II): satisfying local energetic preferences introduces \lq\lq fault zones\rq\rq\ between locally ordered regions.

In our analysis of equatorial atomic distributions (see Fig.~\ref{fig:ordstpic}), we were able to restrict the family of degenerate modes considered to $\mathrm{TEM}_{lm}$ with $l = 0$. To generalize our analysis beyond the equatorial plane, we must consider all modes that meet the degeneracy condition $l + m + n \simeq \sbK R$.
Consider the situation illustrated in Fig.~\ref{fig:frustra}a, first focusing on the non-equatorial sheet marked~(i). Near the center of the sheet, crystallization into modes with $l = 1$ is suppressed because they have low intensity, whereas crystallization into $l = 2$ modes is favored because they have maximal intensity; away from the center, the opposite is true. The change in $l$ forces a change in $m$ or $n$, owing to the degeneracy condition, so the mode functions in the sheet must change across an interfacial zone between the $l = 1$ and $l = 2$ regions. Therefore, either a dislocation, associated with a change in $m$, or an abrupt change in lattice periodicity (i.e., a discommensuration), associated with a change in $n$, is expected~\footnote{This assumes that, as is always the case near threshold, the self-organized lattice is not strong enough to trap the entire atomic distribution at the center or the edge of the sheet.}.

Now consider a situation in which two symmetrically disposed sheets on opposite sides of the equator are populated with atoms, as in sheets~(i) and (ii) of Fig.~\ref{fig:frustra}b.  Atoms in sheet~(i) and those in sheet~(ii) are coupled via the cavity modes. Because $l = 2$ ($l = 1$) mode functions are symmetric (antisymmetric) about the equatorial plane, the atoms in the $l = 2$ ($l = 1$) arrangement in sheet~(ii) occupy the same (opposite) checkerboard as those in sheet~(i). If there are no dislocations, atoms in the interfacial zone remain disordered, because it is impossible for the atoms to satisfy both desiderata (or, equivalently, because the corresponding cavity modes interfere destructively in sheet~(i) and constructively in sheet~(ii)). The introduction of dislocations enables the system to order in part of the interfacial zone, as shown in the right hand side of Fig.~\ref{fig:frustra}b, and is therefore preferred. 

The full many-layer, many-mode system is expected to experience the same kinds of disordering effects as the idealization sketched above: i.e., one expects systems slightly above threshold to develop locally crystalline phases separated by zones riddled with faults. 

\section{Summary and outlook}

Our objective in this paper is to discuss how quantum phases with emergent crystalline order, as well as the associated quantum phase transitions, can be realized and explored using ultracold atomic systems confined in multimode optical cavities. We have shown that the self-organization transition in a concentric cavity realizes Brazovskii's model of stripe formation, formerly confined to soft condensed matter physics, now in a quantal setting, and that the critical fluctuations near this transition are observable in the correlations of the light emitted from the cavity. As in Brazovskii's transition, one expects the weakly self-organized state to be punctuated by defects of unusual morphology, which should be manifest in experiments as dynamically fluctuating domains, and for which there is evidence in simulations \cite{ritsch:personal}. In addition, we have discussed how a multilayered atomic cloud is unable to order globally because of frustration, and is therefore likely to exhibit static domains.

A ubiquitous feature of cavity QED systems, which we hope to elaborate on, is the role played by the delayed response of the cavity on the dynamics of defects and collective excitations. 
This feature is due to the fact that the transient photon population in a high-finesse cavity decays on a timescale proportional to the finesse of the cavity. While critical fluctuations, which take place on extremely long timescales, should not be modified, excitations of the ordered state may be modified by the delayed response. The formalism adopted in this paper can be extended, using influence functional or Keldysh techniques~\cite{fv, keldysh}, to address such issues. However, the slow, long-wavelength nature of the dynamics considered here make such techniques unnecessary for our current purposes. \\

An experimentally relevant question is the extent to which the ``supersolidity'' of BECs in multimode cavities---their simultaneous possession of spontaneous crystalline and superfluid order, discussed here---enables one to test fundamental questions about supersolidity. 
A well-known system in condensed matter that may exhibit supersolidity is solid helium-4; this phenomenon has been studied by ``missing moment of inertia'' experiments~\cite{supersolids}. 
One could imagine performing an analogous experiment using a self-organized BEC in a multimode cavity, and imparting angular momentum to the BEC, e.g., via the pump laser.
Multimode cavity QED provides a tunable setting in which self-organized states spontaneously break the continuous translational and/or orientational symmetries of space, rather than the discrete symmetry between the even and odd sites of either a single mode cavity or an externally imposed lattice.  Thus, it is expected to provide rich and fertile experimental terrain for probing quantum states of matter possessing emergent structural rigidity and superfluidity. This paper explores the bosonic sector of many-body multimode cavity QED; the creation of compliant lattices raises the prospect of realizing fermionic superfluidity via photon-mediated pairing of ultracold fermionic atoms.\\
\\
\begin{acknowledgments}
This work was supported by 
NSF PHY08-47469 (BLL), 
AFOSR FA9550-09-1-0079 (BLL), 
DOE DE-FG02-07ER46453 (PMG), 
and NSF DMR06-05816 (PMG).
\end{acknowledgments}

\end{document}